\documentclass[sigconf, nonacm]{acmart}
\AtBeginDocument{%
  }

\setcopyright{acmlicensed}
\copyrightyear{2025}
\acmYear{2025}
\acmDOI{XXXXXXX.XXXXXXX}
\acmISBN{978-1-4503-XXXX-X/2025/06}

\begin{document}

\title{Beamforming-LLM: What, Where and When Did I Miss?}



\author{Vishal Choudhari}
\email{vc2558@columbia.edu}
\affiliation{%
  \institution{Columbia University}
  \city{New York}
  \state{New York}
  \country{USA}
}

\renewcommand{\shortauthors}{} 

\begin{abstract}
We present Beamforming-LLM, a system that enables users to semantically recall conversations they may have missed in multi-speaker environments. The system combines spatial audio capture using a microphone array with retrieval-augmented generation (RAG) to support natural language queries such as, “What did I miss when I was following the conversation on dogs?” Directional audio streams are separated using beamforming, transcribed with Whisper, and embedded into a vector database using sentence encoders. Upon receiving a user query, semantically relevant segments are retrieved, temporally aligned with non-attended segments, and summarized using a lightweight large language model (GPT-4o-mini). The result is a user-friendly interface that provides contrastive summaries, spatial context, and timestamped audio playback. This work lays the foundation for intelligent auditory memory systems and has broad applications in assistive technology, meeting summarization, and context-aware personal spatial computing.
\end{abstract}




\maketitle

\section{Introduction}
Human auditory attention is inherently limited as we can focus on only one conversation at a time \cite{10.1121/1.1907229}. In social settings like dinner tables, group discussions, or conference poster sessions, we are often surrounded by multiple engaging conversations but end up following just one, inadvertently missing out on others. Similar challenges arise in corporate meetings, where important discussions often unfold in parallel. This highlights a need for a system that can not only capture surrounding conversations but also help users revisit the ones they missed, effectively augmenting the narrow bandwidth of natural auditory attention.

In this work, we introduce Beamforming-LLM, a spatially aware system that captures multi-party conversations, separates them using beamforming, and enables semantic recall through a natural language interface powered by a large language model (LLM). A user can ask questions like, “What did I miss when I was following the conversation on ABC?”, and the system responds with:

\begin{enumerate}
    \item A summary of other conversations that occurred during the same time (\textbf{what}),
    \item The spatial origin of each conversation (\textbf{where}), and
    \item Timestamped audio snippets for playback (\textbf{when}).
\end{enumerate}

This capability opens up promising use cases, including a personal memory assistant, an intelligent meeting summarizer, and a hearing aid companion, making Beamforming-LLM a step toward richer, more accessible conversational experiences in complex auditory environments.

\section{Related Work}

Several recent systems have explored memory in conversational AI, notably Memolet and ConvLogRecaller, which focus on recalling past user interactions to support long-term engagement and personalized assistance \cite{yen2024memolet, lee2024convlogrecaller}. While these approaches demonstrate the value of conversational memory, they primarily operate on past dialogue logs or lifelog data, and are not designed to handle simultaneous multi-speaker environments or spatially distributed audio sources. In parallel, beamforming has been widely used in speech enhancement, particularly for hearing aids, and Retrieval-Augmented Generation (RAG) has proven effective for grounding language model responses in large knowledge bases \cite{doclo2010acoustic, lewis2020retrieval}. However, prior works rarely integrate these technologies into a cohesive system. Our work bridges this gap by combining beamforming for spatial source separation, automatic speech recognition for transcription, and LLM-RAG for semantic querying and summarization, enabling a new class of systems for real-time, spatially aware conversational recall that existing approaches do not support.

\section{Methods}

We assume that each conversation in a multi-speaker environment can be approximated as a spatial point source. Our system is designed to spatially separate these sources, transcribe their content, embed them into a semantic vector space, and retrieve relevant information through a natural language interface. While the individual components leverage well-established techniques, we integrate them into a novel pipeline that enables semantic recall of missed conversations.

\subsection{Beamforming and Directional Source Separation}

Beamforming is a technique widely used in microphone arrays, radar, and wireless communication systems to isolate sound or signal sources based on their spatial origin\cite{adel2012beamforming}. By estimating the Direction of Arrival (DOA) of incoming audio signals, beamforming algorithms apply spatial filters to enhance signals from specific directions while suppressing others.

We use the miniDSP UMA-8 microphone array, which consists of seven microphones—six arranged on a circular perimeter (60° apart) and one at the center, forming a circular array with a diameter of 90 mm. With this configuration, we record multi-channel audio in real-world environments.

Using the Pyroomacoustics library, we estimate directions of arrival (DOAs) and compute \textbf{Minimum Variance Distortionless Response (MVDR)} beamforming filters to spatially isolate conversations\cite{scheibler2018pyroomacoustics, habets2010mvdr}. The output is a set of directionally filtered \texttt{.wav} files, each representing a distinct spatial conversation stream, e.g., \texttt{\{`left': abc.wav, `right': def.wav, `front': ghi.wav\}}.

\subsection{Automatic Speech Recognition (ASR)}
To support semantic querying, the directionally separated audio must be transcribed. We use Whisper, an open-source ASR model from OpenAI trained on over 680,000 hours of multilingual speech. Whisper is chosen for its robustness across accents, speaking styles, and noisy environments \cite{radford2023robust}.

Transcribing each beamformed audio stream yields timestamped text segments, which are chunked into semantically meaningful units. These segments form the input to the retrieval pipeline and also allow precise alignment with the original audio.

\subsection{Retrieval-Augmented Generation with Vector Embeddings}

Since users may query information spanning hours of audio, we adopt a Retrieval-Augmented Generation (RAG) architecture to overcome the limited context window of current LLMs\cite{gao2023retrieval}. Each conversation transcript is divided into chunks of approximately three sentences, and embeddings are computed using the MiniLM sentence encoder.

These embeddings are indexed in FAISS, an efficient, GPU-scalable vector database that supports fast approximate nearest neighbor search \cite{douze2024faiss}. Each chunk is assigned a unique index and stored along with metadata including:

\begin{itemize}
    \item Chunk text (\textbf{what})
    \item Direction of arrival (\textbf{where})
    \item Start and end timestamps (\textbf{when})
\end{itemize}

This metadata is maintained in a separate dictionary to support timestamp-based lookups and audio playback.

\subsection{Natural Language Interface and Semantic Retrieval}

The system supports natural language queries such as:  
\begin{quote}
\emph{``What did I miss when I was following the conversation about dogs?''}
\end{quote}

To process such queries:

\begin{enumerate}
    \item An LLM extracts the \textbf{topic} (e.g., \texttt{"dogs"}) from the user’s question.
    \item The topic is embedded using the same encoder as the database and used to query FAISS for the \textbf{top-10 semantically similar chunks} from the attended conversation stream.
    \item Retrieved chunks are further \textbf{filtered for true relevance} using GPT-4o-mini.
    \item These filtered results act as \textbf{centroids}, and additional nearby chunks (within a window size $K$) are grouped to form extended conversational \textbf{snippets}.
\end{enumerate}

The metadata for these snippet chunks is then used to:
\begin{itemize}
    \item Identify \textbf{where} the attended conversation took place
    \item Retrieve \textbf{temporally overlapping} non-attended conversations
    \item Summarize the missed conversations using the LLM in a bullet-point format like: \\
    \emph{``While you were listening to X, you missed Y.''}
\end{itemize}

\begin{figure*}[t!]
  \centering
  \includegraphics[width=0.5\textwidth]{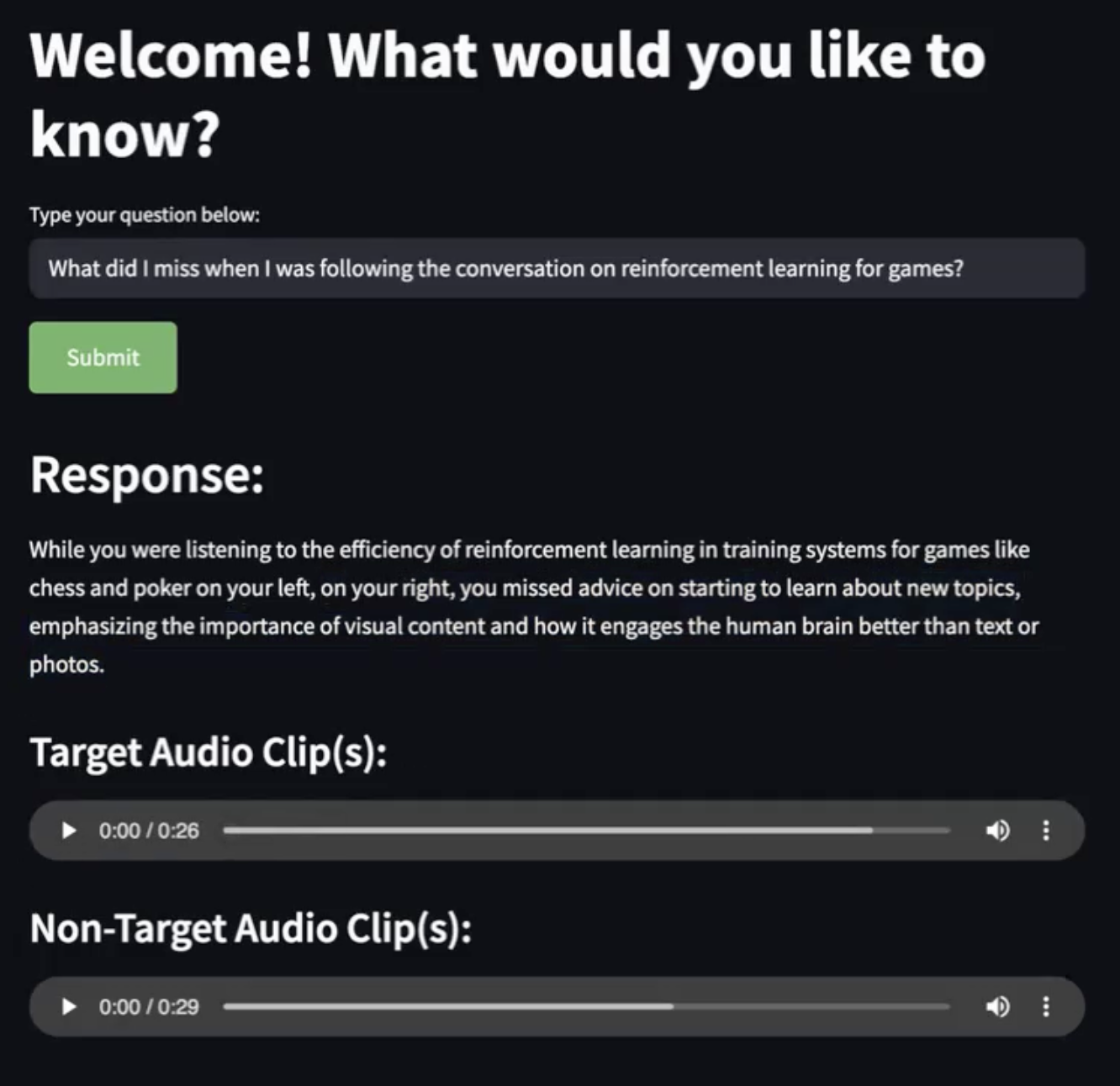} 
  \caption{Graphical user interface of Beamforming-LLM. Users enter natural language queries to recall missed conversations. The system generates a contrastive summary and provides timestamped audio snippets from both the attended and unattended conversation streams.}
  \label{fig:gui}
\end{figure*}

Since all segments are timestamped, users can \textbf{replay relevant audio snippets} for both attended and missed conversations, enabling seamless auditory recall. The system GUI is shown in Figure \ref{fig:gui}.

\subsection{Deployment and LLM Choice}
\begin{figure*}[t!]
  \centering
  \includegraphics[width=\textwidth]{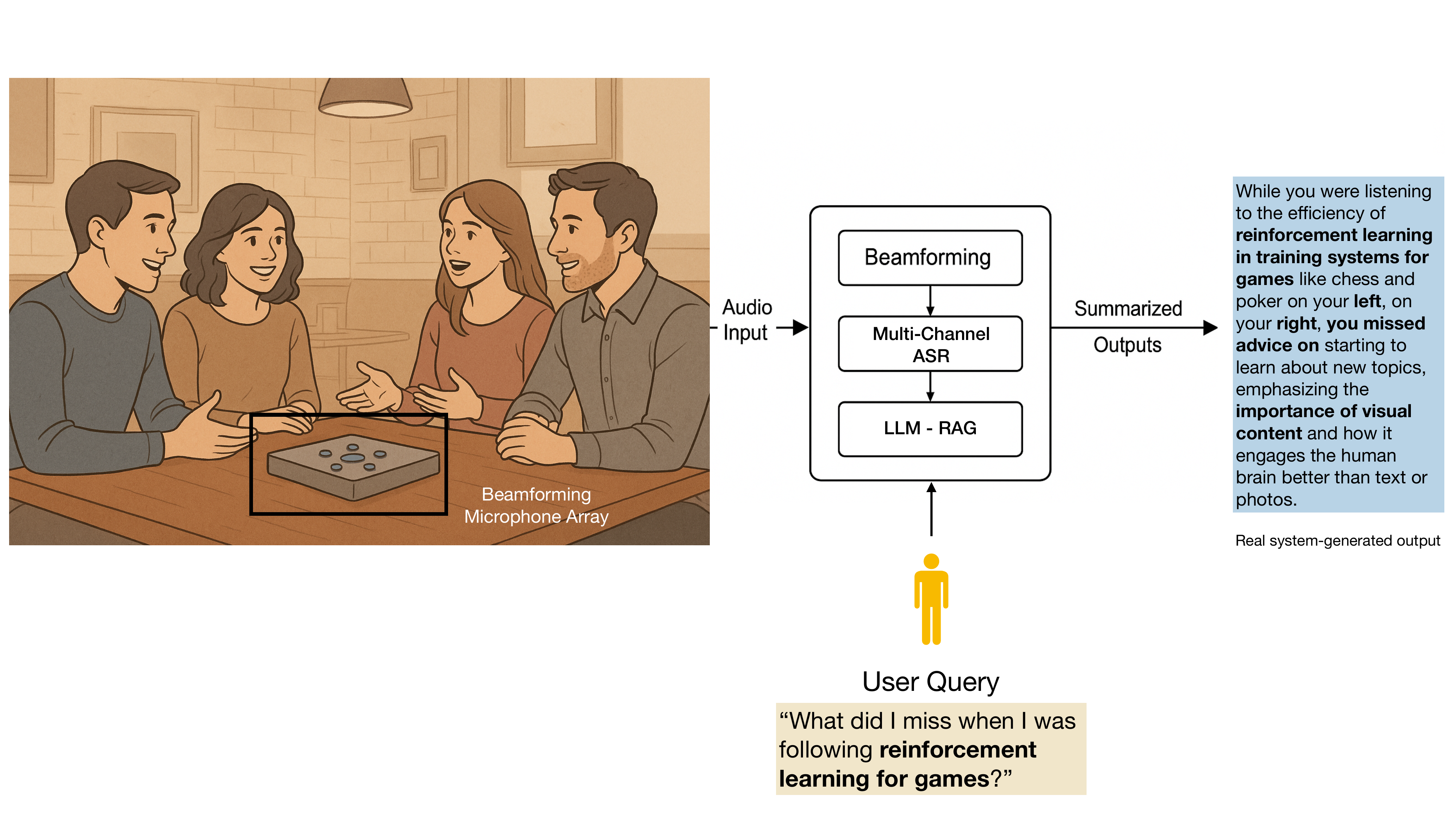} 
  \caption{Conceptual overview of Beamforming-LLM. (Left) A tabletop beamforming microphone array captures multi-party conversations in a social setting. (Center) The system pipeline consists of three stages: spatial filtering via beamforming to separate multiple audio streams, multi-channel transcription via automatic speech recognition (ASR), and semantic summarization via a large language model (LLM) upon receiving a user query. (Right) The output provides directional summaries of distinct conversations, enabling users to recall what was said from different parts of the environment.}
  \label{fig:system-overview}
\end{figure*}

The entire pipeline can be deployed on resource-constrained edge devices such as a Raspberry Pi, making it feasible for real-time and mobile applications. We use OpenAI’s GPT-4o-mini as our LLM of choice for both query understanding and summarization. To keep the system lightweight, we access the model via API calls, offloading heavy computation to the cloud. The system framework is shown in Figure \ref{fig:system-overview}.

\section{Results}

\begin{figure}[htbp]
  \centering
  \includegraphics[width=\columnwidth]{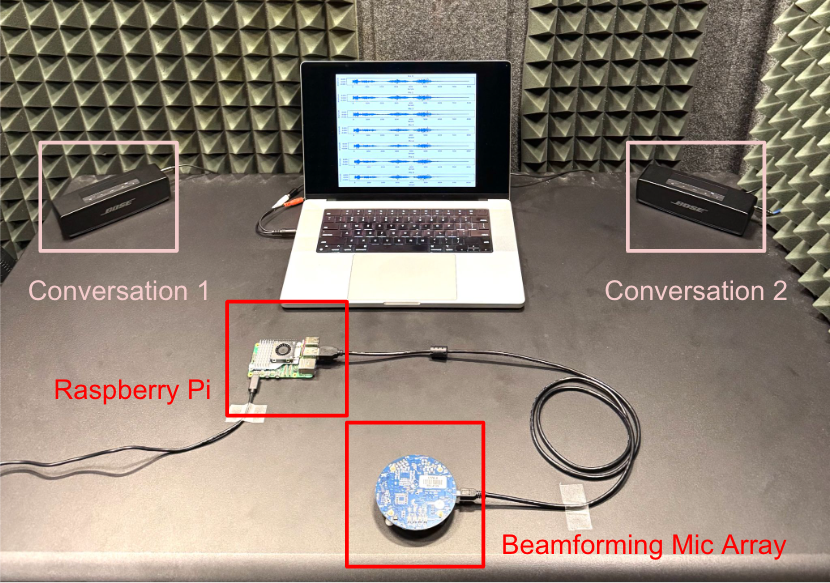}
  \caption{Experimental setup for evaluating Beamforming-LLM. A laptop plays two simultaneous conversations through separate left and right speakers to simulate spatially distinct discussions. The beamforming microphone array (miniDSP UMA-8), connected to a Raspberry Pi, is positioned at the listener’s location to capture the multi-channel audio for spatial separation and semantic recall.}
  \label{fig:testSetup}
\end{figure}

\begin{figure*}[t!]
  \centering
  \begin{subfigure}[t]{0.32\textwidth}
    \centering
     \includegraphics[height=4.5cm]{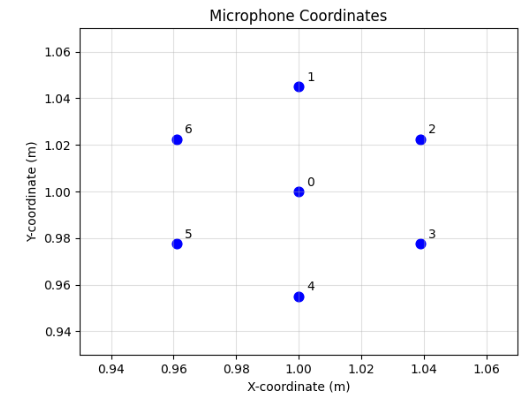}
    \caption{Mic array geometry}
    \label{fig:subfig-a}
  \end{subfigure}
  \hfill
  \begin{subfigure}[t]{0.32\textwidth}
    \centering
     \includegraphics[height=4.5cm]{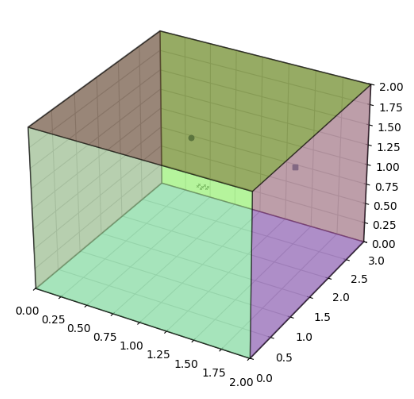}
    \caption{Room configuration}
    \label{fig:subfig-b}
  \end{subfigure}
  \hfill
  \begin{subfigure}[t]{0.32\textwidth}
    \centering
     \includegraphics[height=4.5cm]{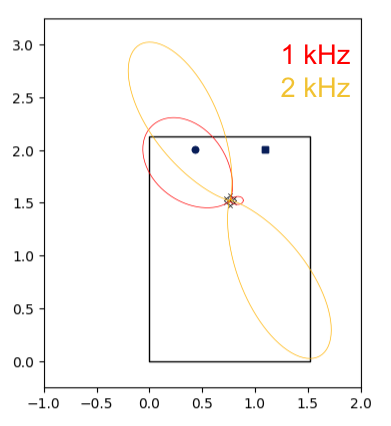}
    \caption{Beam patterns}
    \label{fig:subfig-c}
  \end{subfigure}
  \caption{
  Microphone array configuration and beamforming. (a) Geometry of the miniDSP UMA-8 circular microphone array with seven microphones, labeled 0–6. (b) Simulated 3D room setting used for direction of arrival (DOA) estimation and beamforming filter computation. (c) Beam patterns visualized at 1 kHz (red) and 2 kHz (orange), demonstrating the array’s directional selectivity for isolating spatially distinct sound sources.}
  \label{fig:beamforming}
\end{figure*}

To evaluate our system end-to-end, we conducted a controlled tabletop experiment using the setup illustrated in Figure \ref{fig:testSetup}. A Raspberry Pi connected to a miniDSP UMA-8 microphone array was placed at the center of the lower edge of a rectangular table, simulating a listener’s position. At the opposite side, a laptop was positioned at the center of the top edge, with its left and right speakers playing two simultaneous podcast conversations. The left speaker played a podcast on artificial intelligence featuring Yann LeCun, while the right speaker played a podcast on personal finance and wealth management. This arrangement was designed to mimic natural conversational clusters at a dinner table or social gathering. In total, we recorded 60 minutes of multi-conversation audio.

\subsection{Beamforming Performance}

We applied beamforming to the recorded multi-channel audio to spatially isolate the two sources. Using Pyroomacoustics, we simulated the room geometry and microphone layout to estimate direction of arrival (DOA) and compute MVDR beamforming filters. The spatial separation enabled by beamforming is illustrated in Figure \ref{fig:beamforming}, which shows the estimated room setup and frequency-dependent beam patterns. 

To quantify the enhancement achieved through beamforming, we measured two standard speech quality metrics:
\begin{enumerate}
    \item PESQ (Perceptual Evaluation of Speech Quality) \cite{rix2001perceptual}
    \item STOI (Short-Time Objective Intelligibility) \cite{taal2010short}
\end{enumerate}

Results showed consistent improvement in both metrics for each source, as shown in Table \ref{tab:stoi_pesq}. These improvements confirm that beamforming significantly enhanced the clarity and intelligibility of each spatial source, laying a reliable foundation for downstream transcription and retrieval.

\begin{table}[h]
\centering
\caption{STOI and PESQ scores before and after beamforming for left and right conversations.}
\begin{tabular}{lccc}
\toprule
\textbf{Metric} & \textbf{Before} & \textbf{After} & \textbf{Improvement} \\
\midrule
\multicolumn{4}{l}{\textit{Left Conversation}} \\
\hspace{1em}STOI & 0.23 & 0.63 & +0.41 \\
\hspace{1em}PESQ & 1.35 & 2.50 & +1.15 \\
\midrule
\multicolumn{4}{l}{\textit{Right Conversation}} \\
\hspace{1em}STOI & 0.33 & 0.70 & +0.38 \\
\hspace{1em}PESQ & 1.46 & 2.35 & +0.88 \\
\bottomrule
\end{tabular}
\label{tab:stoi_pesq}
\end{table}

\subsection{Retrieval Pipeline Evaluation}
The semantic retrieval and summarization pipeline was tested using queries related to both conversations. The system was able to accurately return relevant segments in response to prompts such as ``What did I miss when I was listening to the AI conversation?'' and ``Summarize what was being discussed on the other side during the finance talk.'' The generated summaries correctly identified key topics and provided natural, contrastive phrasing.

While these early results are promising, a more comprehensive evaluation will involve deploying the system to a broader set of users and collecting qualitative feedback and quantitative measures of satisfaction, accuracy, and usability. Conducting user studies in real-world multi-speaker environments is planned as future work.

\section{Discussion and Conclusion}

This work presents a functional MVP for spatially grounded conversational recall, but several limitations remain. The current setup depends on a controlled form factor and line-of-sight audio capture, which may not generalize well to complex environments. Beamforming alone struggles with overlapping speakers and off-axis sources, suggesting the need to integrate speech separation and enhancement techniques in future iterations \cite{luo2019conv}. The system currently treats conversations as 2D point sources; extending to 3D spatial localization would improve applicability in vertical spaces like classrooms or auditoriums. Other areas for technical improvement include handling cross-talk between conversation clusters, incorporating speaker diarization (e.g., via Pyannote-Whisper), and generalizing beyond conversations to capture other acoustic events \cite{bredin2020pyannote}.

Beyond technical extensions, the system offers exciting opportunities for multimodal expansion. A voice-based interface could replace typed queries, and integrating attention signals (e.g., gaze or EEG) would enable personalized recall of what users likely missed \cite{mesgarani2012selective, jiang2025aad}. Geotagging and vision-based cues could further contextualize spatial summaries, while ethical safeguards such as consent and on-device processing must be prioritized. Ultimately, this platform could evolve into a personal memory assistant, intelligent meeting summarizer, or even a hearing aid companion, amplifying human attention in dynamic auditory environments.

\section{Acknowledgements}
The author would like to thank Dr. Xiaofan Jiang, Dr. Parijat Dubey, and Dr. Chen Wang for their valuable guidance and discussions. The author is also grateful to Swethaashri Ramesh and Kirtana Kalidindi for their helpful feedback and support during the development of this work.

\bibliographystyle{ACM-Reference-Format}
\bibliography{myReferences}

%








\end{document}